\documentstyle[aps,pra,twocolumn,epsfig,floats]{revtex}

\def\Red  {}
\def\Black{}
\def\Blue {}
\begin{document}
\draft
\title{\Red Quantum theory of self-action of ultrashort light pulses\\
       in an inertial nonlinear medium\Black }
\author{F. Popescu${}^{*}$\and A. S. Chirkin${}^{\dagger}$}
\address{Moscow State University,\\
         119899 Moscow, Russian Federation\\    [1mm]
        {\sf E-mail: ${}^{*}$florentin\_p@hotmail.com,
                ${}^{\dagger}$chirkin@foton.ilc.msu.su}}
\date{April 10, 2000}
\twocolumn[
\widetext
\begin{@twocolumnfalse}
\maketitle
\begin{abstract}\Blue
The systematic theory of the formation of the short light pulses in the squeezed
state during the propagation in a medium with inertial Kerr nonlinearity is
developed. The algebra of time-dependent Bose-operators is elaborated and the
normal-ordering theorem for them is formulated. It is established that the spectral
region where the quadrature fluctuations are weaker than the shot-noise, depends on
both the relaxation time of the nonlinearity and the magnitude of the nonlinear phase
shift. It is also shown that the frequency at which suppression of the fluctuation is
greatest can be controlled by adjusting the phase of the initial coherent light
pulse. The spectral correlation function of photons is introduced and photon
antibunching is found.\Black 
\end{abstract}
\vspace{0.5cm}
\narrowtext
\end{@twocolumnfalse}]
\section{INTRODUCTION}
During the past years the study of formation of nonclassical light pulses  in
nonlinear media has been the focus of a considerable attention. The present article
is devoted to the development of the consecutive theory of formation of nonclassical
short light pulses in nonlinear media with inertial Kerr nonlinearity. It is well
known, that in nonlinear media in the presence of the self-action effect the
squeezing of quantum fluctuations of one quadrature component of a field with
conservation of the photon statistics take place \cite{Kitagava}. At present there
are two basic directions of research in the quantum theory of self-action of
ultrashort light pulses (USPs). In the first approach
\cite{Hashiura,Nishizawa,Shirashi,Ahmanov} the calculations of the nonclassical light
formation at the self-action of light pulses assume that the nonlinear response of
nonlinearity is instantaneous and that the relative fluctuations are small. This
latter assumption is valid for the intensive USPs frequently used in experiments.
However, a finite relaxation time of the nonlinearity has a principled importance as
the relaxation time determines the region of the spectrum of the quantum fluctuation
below the standard noise level. For the first time, in \cite{Blow} was noted that for
the correct quantum solution of self-action, it is necessary to take into account the
presence of quantum noise. The presence of quantum noise  was anticipated in
\cite{Boivin} as thermal addition to the interaction Hamiltonian. This addition was
necessary in order to satisfy the commutation relation for time-dependent
Bose-operators. If the interaction Hamiltonian has the normally ordered form
\cite{Popescu} then  it is not necessary to have deal with thermal fluctuations. This
approach allows us to develop an algebra of time-dependent Bose-operators and to
investigate the spectrum of quantum fluctuations of the quadrature components. The
results of the quantum theory of self-action for USPs in the medium with the
relaxation Kerr nonlinearity based on the normally ordered interaction Hamiltonian
and the developed algebra of time-dependent Bose-operators are presented below.
\section{THE QUANTUM EQUATION OF SELF-ACTION OF USP{\scriptsize{s}}}
For the monochromatic radiation the quantum equation of self-action can be found, for
example, in \cite{Ahmanov}. Up to now, the proper quantum equation of the self-action
for USPs with the account  of relaxation behavior of the nonlinearity is absent in
the literature. It is necessary to mention that the deduction of the quantum equation
is based on the interaction Hamiltonian. However, in this case we obtain the
time-evolution equation for the Bose-operators. For monochromatic radiation the
conversion of time-evolution equation into space-evolution one is done using the
$t\smash{\longrightarrow}z/u$ replacement, where $z$ is time variable and $u$ is the
speed of pulse in nonlinear media. If we deal with the propagation of pulse in
nonlinear media then the "impulse operator" of a pulse field should be used
\cite{Mooki}. We begin with the analyse of the self-action of UPSs in non-inertial
nonlinear media.
\subsection{THE QUANTUM EQUATION OF SELF-ACTION IN THE NON-INERTIAL NONLINEAR
MEDIA} In nonlinear media with non-inertial behavior the self-action process is
described using the impulse operator (quantity of movement) $\hat{G}_{int}(z)$
\cite{Boivin}
\begin{equation}\label{gam}
\hat{G}_{int}(z)=\beta\hbar
\!\!\int_{-\infty}^{\infty}\!\!\hat{\mathbf{N}}[\hat{n}^{2}(t,z)]dt,
\end{equation}
where $\hat{\mathbf{N}}$ is the operator of normal ordering, factor $\beta$ is
defined by the cubic nonlinearity of the medium \cite{Ahmanov}. In consequence, in
the Heisenberg representation the quantum space-evolution equation for the
annihilation photons Bose-operator in a given cross section $z$ ($\hat{A}(t,z)$) has
the form
\begin{equation}\label{a1}
-i\hbar\frac{\partial\hat{A}(t,z)}{\partial
z}=\left[\hat{A}(t,z),\hat{G}_{int}(z)\right].
\end{equation}
In agreement with (\ref{gam}) the quantum equation of self-action for a light pulse
follows from (\ref{a1})
\begin{equation}\label{Eq}
\frac{\partial\hat{A}(t,z)}{\partial z}-i\beta\hat{A}^{+}(t,z)\hat{A}^{2}(t,z)=0.
\end{equation}
Eq.(\ref{Eq}) is written in the moving coordinate system: $z\smash{=}z'$ and
$t\smash{=}t'\smash{-}z/u$. It is important to note that in comparison with the
so-called nonlinear Heisenberg equation, used in the quantum theory of optical
solitons, in (\ref{Eq}) the dissipation of light pulse in the nonlinearity is not
taken into account. This approach corresponds to the first approximation of
dissipation theory. In fact, the traditional way to introduce the quantum equation of
self-action is based on the interaction Hamiltonian. In this case one gets the
time-evolution equation. The transition to the space-evolution, as already mentioned,
is realized using replacement $t\longrightarrow z/u$. This approach is enough
reasonable in case the radiation is monochromatic. If we deal with the nonlinear
propagation of a pulse, we use the impulse operator of a pulse field (\ref{gam})
which is connected with the evolution of field in space. Eq.(\ref{Eq}) has the
solution
\begin{equation}\label{clas1}
\hat{A}(t,z)=e^{i\gamma\hat{n}_0(t)}\hat{A}_0(t),
\end{equation}
where $\gamma\smash{=}\beta z$ and, as usual, $\hat{A}_0(t)$ is the value of the
operator at input of nonlinear media
($\hat{A}_{0}(t)\smash{=}\hat{A}(t,z\smash{=}0)$),
$\hat{n}_{0}(t)\smash{=}\hat{A}^{+}_{0}(t)\hat{A}_{0}(t)$ is the photon number
``density'' operator. For its hermitian conjugated operator we find
\begin{equation}\label{clas2}
\hat{A}^{+}(t,z)=\hat{A}^{+}_0(t)e^{-i\gamma\hat{n}_0(t)}.
\end{equation}
In agreement with (\ref{clas1}) and (\ref{clas2}) the operator
$\hat{n}(t,z)=\hat{A}^{+}(t,z)\hat{A}(t,z)$ does not change itself in nonlinear
medium:
\begin{equation}\label{nor1}
\hat{n}(t,z)=\hat{n}(t,z\smash{=}0)=\hat{n}_0(t),
\end{equation}
where $z\smash{=}0$ corresponds to the input of the nonlinear medium. In fact,
(\ref{nor1}) means that the photon statistic in media remains unchanged. The
commutation relation at the input ($z\smash{=}0$) of nonlinearity
$[\hat{A}_{0}(t_{1}),\hat{A}^{+}_{0}(t_{2})]=\delta(t_{1}\smash{-}t_{2})$, should be
satisfied for any coordinate $z$ in nonlinear media:
\begin{equation}\label{delta2}
[\hat{A}(t_{1},z),\hat{A}^{+}(t_{2},z)]=\delta(t_{1}-t_{2}).
\end{equation}
The solutions (\ref{clas1}) and (\ref{clas2}) do not permit to verify the commutation
relation (\ref{delta2}). Besides, the analyse of the statistical characteristics of
the pulse is accompanied by the necessity of the reduction to the normally ordered
form of the expression $e^{i\gamma\hat{n}_0(t)}$. In this case, the solutions
(\ref{clas1}) and (\ref{clas2}) are accompanied by the singularity of the function
$\delta(t)$ at $t\smash{=}0$. The specified circumstances represent the main
deficiency of the quantum theory of self-action of USPs in non-inertial nonlinear
media.
\subsection{THE QUANTUM EQUATION OF SELF-ACTION IN THE INERTIAL NONLINEAR
MEDIA} In the classical theory, the self-action process in inertial nonlinear media
is described by the equation (in the first approximation of the dispersion theory)
\cite{Ahmanov}
\begin{eqnarray}\label{pirmi}
\frac{\partial B(t,z)}{\partial z}+\frac{1}{u}\frac{\partial B(t,z)}{\partial t}
-i\frac{k_0}{n_0}\Delta n(|&B&(t,z)|^2)\nonumber\\ &{}&\times B(t,z)=0,
\end{eqnarray}
where: $B(t,z)$- the complex amplitude of a pulse, $z$- the distance in the nonlinear
media, $u$- the group velocity, $\Delta n(|B(t,z)|^2)\smash{=}\Delta n((t,z)$- the
nonlinear addition to the coefficient of refraction. We consider that the last term
of (\ref{pirmi}) is caused by the high-frequency Kerr effect, and its evolution
follows from the equation
\begin{equation}\label{army}
\tau_r\frac{\partial\Delta n(t,z)}{\partial t}+\Delta
n(t,z)=\frac{1}{2}\,n_2|B(t,z)|^2.
\end{equation}
Here $\tau_r$ represents the relaxation time of the nonlinearity and $n_2$- the
nonlinear factor. We mention that in general the behaviour of the nonlinear addition
differs from the one characterized by (\ref{army}). However, if the carrying
frequency of a pulse is far enough from the resonance and the pulse duration $\tau_p$
is greater that the relaxation time $\tau_r$, then (\ref{army}) is correct
\cite{Ahmanov}. The solution of (\ref{army}) looks like
\begin{eqnarray}\label{masa}
\Delta n(t,z)=(n_2/2)\int_{-\infty}^{t}H(t-t_1)|B(t_1,z)|^2\,dt_1.
\end{eqnarray}
The function of nonlinear response $H(t)$ is entered in such a way that in the limit
$\tau_p\smash{\gg}\tau_r$ the nonlinear addition becomes $\Delta
n(t,z)\smash{=}(n_2/2)|B(t,z)|^2$. In the moving system of coordinates
$(t'\smash{=}t\smash{-}z/u,z'\smash{=}z)$, taking into account (\ref{masa}),
eq. (\ref{pirmi}) takes the form
\begin{equation}\label{tata}
\frac{\partial B(t,z)}{\partial
z}=i\gamma^{\star}\int_{0}^{\infty}H(t_1)|B(t-t_1,z)|^2 \, dt_1B(t,z),
\end{equation}
where $\gamma^{\star}\smash{=}k_0n_2/2n_0$ and the quotation-marks (~${'}$~) in new
system of coordinates further will be lowered for simplicity. The transition to the
quantum equation usually is carried out in the spectral representation. However, in
the considered case it is more natural to use time representation. We make in
(\ref{tata}) replacement of the complex amplitudes with the operators, entered in the
previous sections,
\begin{equation}\label{mama}
B(t,z)\rightarrow iC\hat{A}(t,z),\quad B^{*}(t,z)\rightarrow -iC\hat{A}^{+}(t,z).
\end{equation}
The right part of the equation we have gotten in this way, will be written below in
normally ordered form. In order to take into account the presence of the vacuum
fluctuations, existing up to the moment of arrival of the pulse, we will replace in
(\ref{tata}) the bottom limit of integration by ${-\infty}$. As a result we get
($C\smash{=}(\hbar\omega_{0}/2V)^{1/2}$, $\beta\smash{=}\gamma^{\star}C^{2}/2$)
\begin{eqnarray}\label{sora}
\frac{\partial\hat{A}(t,z)}{\partial z}=i\beta\int_{-\infty}^{\infty}
H(|t_1|)\hat{A}^{+}(t\smash{-}t_1,z)&{}&\hat{A}(t\smash{-}t_1,z)\nonumber\\
&{}&\times\,\hat{A}(t,z)dt_1.
\end{eqnarray}
Eq.(\ref{sora}) represents the correct quantum equation and can be obtained from
space evolution equation for the operator $\hat{A}(t,z)$ in interaction
representation (\ref{a1}). Taking into consideration the inertial behaviour of the
nonlinearity, the impulse operator of a pulse should be introduced as:
\begin{eqnarray}\label{ham}
\hat{G}_{int}(z)=\hbar\beta\int_{-\infty}^{\infty}dt&{}&\int_
{-\infty}^{t}H(t-t_{1})\nonumber\\
&{}&\times\hat{\mathbf{N}}\left[\,\hat{n}(t,z)\hat{n}(t_{1},z)\right]dt_{1},
\end{eqnarray}
where $H(t)$ is the nonlinear response of the medium (see (\ref{masa}))
($H(t)\smash{\neq}0$ at $t\smash{\geq}0$ and $H(t)\smash{=}0$ at $t\smash{<}0$). We
note that the integral expression in (\ref{ham}) at the moment of time $t$ depends
only on the previous ones. Therefore, in this case the causality principle relative
to measured physical value is not broken. An important condition that must be
satisfied by the impulse operator is
\begin{equation}
[\hat{G}_{int}(z),\hat{n}(t,z)]=[\hat{G}_{int}(z),\hat{n}_0(t)]=0,
\end{equation}
which means that the photon number operator remains unchanged in medium (photon
number is a constant of motion \cite{Kitagava}). The annihilation Bose-operator which
agrees with the (\ref{ham}) verifies the equation of self-action
\begin{equation}\label{eqe}
\frac{\partial\hat{A}(t,z)}{\partial z}-i\beta q[\hat{n}_{0}(t)]\hat{A}(t,z)=0,
\end{equation}
where
\begin{equation}\label{quat}
q[\hat{n}_{0}(t)]=\int_{0}^{\infty}H(t_{1})\left[\,\hat{n}_{0}(t-t_{1})
+\hat{n}_{0}(t+t_{1})\right]dt_{1}.
\end{equation}
If $H(t)\smash{=}\delta(t)$ then (\ref{eqe}) is converted into (\ref{Eq}).
Eq.(\ref{Eq}) can be obtained in the same form if we consider that the response of
the nonlinearity has relaxation behaviour $H(t)$ in accordance with (\ref{masa}) and
the relaxation time $\tau_{r}$ is significantly less than the duration of a pulse
$\tau_{p}$. At the same time, as will be shown further, in this limited case also,
the account of finite relaxation time plays the main role in formation of the
nonclassical light. It is necessary to note, that by replacement in (\ref{eqe}) the
operators on complex values we obtain the equation, which not completely coincides
with the classical equation of self-action in presence of the non-stationary
nonlinear response. There is no second term, included in (\ref{quat}) as
$\hat{n}_{0}(t+t_{1})$. The presence of this term in quantum theory, which at the
first sight is in contradiction with the causality principle, is connected with the
quantum description that even in absence of a pulse the vacuum fluctuations always
are present. A response function was introduced already in the model for the Kerr
effect by Blow et al. \cite{Blow}. Although the need for an attendant noise source
was anticipated by these authors, they did not indicate where it should be inserted.
In \cite{Boivin} the quantum noise as thermal fluctuations was additive inserted in
the interaction Hamiltonian but this procedure did not allow to develop the
consistent quantum theory of self-action of USPs. According to (\ref{ham}), the
operator $\hat{n}(t,z)$ remains unchanged in nonlinear media (see (\ref{nor1})).
Solving the spatial evolution equation (\ref{eqe}), the annihilation (creation)
photons Bose-operators in nonlinear medium have the following form:
\begin{eqnarray}
\hat{A}(t,z)&=&e^{i\gamma q[\hat{n}_{0}(t)]}\,\hat{A}_{0}(t).\label{A}\\
\hat{A}^{+}(t,z)&=&\hat{A}^{+}_{0}(t)\,e^{-i\gamma q[\hat{n}_{0}(t)]}.\label{B}
\end{eqnarray}
In (\ref{A}) and (\ref{B}) $\hat{A}_{0}(t)\smash{=}\hat{A}(t,0)$,
$\gamma\smash{=}\beta z$. The expression $q[\hat{n}_{0}(t)]$ (see (\ref{quat})) it is
convenient to be written as:
\begin{equation}\label{sun}
q[\hat{n}_{0}(t)]=\int_{-\infty}^{\infty}\!\!h(t_{1})\hat{n}_{0}(t-t_{1})
dt_{1},\quad(h(t)=H(|t|).
\end{equation}
If we consider $\hat{n}$ to be time-independent, then the expressions (\ref{A}) and
(\ref{B}) give results for the monochromatic field. In the case that the nonlinear
response in (\ref{A}), (\ref{B}) has the form $h(t)=\delta(t)$, the results for
non-inertial nonlinear media can be obtained (see (\ref{clas2})). To find the
statistical characteristics of a pulse at the output of the nonlinear medium it is
necessary to estimate the averages of the operators $\hat{A}(t,z)$,
$\hat{A}^{+}(t,z)$ and their combinations. They can be estimated, if the operator
expressions are given in the normally ordered form, when the creation photons
Bose-operators are placed at the left of the annihilation photons Bose-operators. The
use of the expressions (\ref{A}), (\ref{B}) involves the development of a special
mathematical device.
\section{THE ALGEBRA OF THE TIME-DEPENDENT BOSE-OPERATORS}
For the beginning, in order to simplify some expressions we introduce the operators:
\begin{equation}\label{oul}
\hat{O}(t)=i\gamma q[\hat{n}_{0}(t)],\qquad
\hat{O}^{+}(t)=-i\gamma q[\hat{n}_{0}(t)],
\end{equation}
where $\hat{O}^{+}(t)\smash{=}-\hat{O}(t)$. Hence, the equations of self-action
(\ref{A}), (\ref{B}) can be represented as:
\begin{eqnarray}
\hat{A}(t,z)&=&e^{\hat{O}(t)}\,\hat{A}_{0}(t),\label{sta1}\\
\hat{A}^{+}(t,z)&=&\hat{A}^{+}_{0}(t)\,e^{\hat{O}^{+}(t)}.\label{sta2}
\end{eqnarray}
Taking into account (\ref{sun}), is it easy to remark that
\begin{equation}
[\hat{O}(t_{1}),\hat{O}(t_{2})]=0.
\end{equation}
In consequence we have
\begin{eqnarray}
e^{\hat{O}(t_{1})}e^{\hat{O}(t_{2})}&=&e^{\hat{O}(t_{2})}e^{\hat{O}(t_{1})}\nonumber\\
&=&e^{\hat{O}(t_{1})+\hat{O}(t_{2})}=e^{\hat{O}(t_{2})+\hat{O}(t_{1})}.
\end{eqnarray}
\subsection{THE PERMUTATION OPERATOR RELATIONS} The following operator permutation
relations hold:
\begin{eqnarray}
\hat{A}_{0}(t_{1})\hat{O}(t_{2})&=&[\hat{O}(t_{2})+i\gamma
h(t_{2}-t_{1})]\hat{A}_{0}(t_{1}),\nonumber\\
\hat{A}_{0}(t_{1})\hat{{O}^{+}}(t_{2})&=&[\hat{O}^{+}(t_{2})-i\gamma
h(t_{2}-t_{1})]\hat{A}_{0}(t_{1}),\nonumber\\
\hat{A}^{+}_{0}(t_{1})\hat{O}(t_{2})&=&[\hat{O}(t_{2})-i\gamma
h(t_{2}-t_{1})]\hat{A}^{+}_{0}(t_{1}),\\
\hat{A}^{+}_{0}(t_{1})\hat{{O}^{+}}(t_{2})&=&[\hat{O}^{+}(t_{2})+i\gamma
h(t_{2}-t_{1})]\hat{A}^{+}_{0}(t_{1}).\nonumber
\end{eqnarray}
Using the mathematical induction it is possible to demonstrate the validity of the
formulae ($m\in N$):
\begin{eqnarray}
\hat{A}_{0}(t_{1})\hat{O}^{m}(t_{2})&=&[\hat{O}(t_{2})+i\gamma
h(t_{2}-t_{1})]^{m}\hat{A}_{0}(t_{1}),\nonumber\\
\hat{A}_{0}(t_{1})\hat{{O}^{+}}^{m}(t_{2})&=&[\hat{O}^{+}(t_{2})-i\gamma
h(t_{2}-t_{1})]^{m}\hat{A}_{0}(t_{1}),\nonumber\\
\hat{A}^{+}_{0}(t_{1})\hat{O}^{m}(t_{2})&=&[\hat{O}(t_{2})-i\gamma
h(t_{2}-t_{1})]^{m}\hat{A}^{+}_{0}(t_{1}),\\
\hat{A}^{+}_{0}(t_{1})\hat{{O}^{+}}^{m}(t_{2})&=&[\hat{O}^{+}(t_{2})+i\gamma
h(t_{2}-t_{1})]^{m}\hat{A}^{+}_{0}(t_{1}).\nonumber
\end{eqnarray}
To simplify the operator algebra is useful to redefine
\begin{equation}
\emph{G}(t_{2}\smash{-}t_{1})\smash{=}i\gamma h(t_{2}\smash{-}t_{1}),\quad
\emph{G}^{*}(t_{2}\smash{-}t_{1})\smash{=}-i\gamma h(t_{2}\smash{-}t_{1}).
\end{equation}
Decomposing $e^{\hat{O}(t)}$ and $e^{\hat{O}^{+}(t)}$ in Taylor series we obtain the
operator permutation relations which play an important role at the estimation of the
statistical characteristics of a pulse. Hence, finally we have:
\begin{eqnarray}
\hat{A}_{0}(t_{1})e^{\hat{O}(t_{2})}&=&e^{\hat{O}(t_{2})+
\emph{G}(t_{2}-t_{1})}\hat{A}_{0}(t_{1}),\nonumber\\
\hat{A}_{0}(t_{1})e^{\hat{O}^{+}(t_{2})}&=&e^{\hat{O}^{+}(t_{2})
+\emph{G}^{*}(t_{2}-t_{1})}\hat{A}_{0}(t_{1}),\nonumber \\
e^{\hat{O}(t_{1})}\hat{A}^{+}_{0}(t_{2})&=&\hat{A}^{+}_{0}(t_{2})
e^{\hat{O}(t_{1})+\emph{G}(t_{2}-t_{1})},\label{Tab}\\
e^{\hat{O}^{+}(t_{1})}\hat{A}^{+}_{0}(t_{2})&=&\hat{A}^{+}_{0}(t_{2})
e^{\hat{O}^{+}(t_{1})+\emph{G}^{*}(t_{2}-t_{1})}.\nonumber
\end{eqnarray}
Using the permutation relations (\ref{Tab}) it is possible to verify the commutation
relation (\ref{delta2}) for the operators $\hat{A}(t,z)$ and $\hat{A}^{+}(t,z)$.
\subsection{THE NORMAL ORDERING THEOREM} As pointed out previously, in the
considered analyse another important question is represented by the reduction to
normally ordered form of the operators $\hat{A}(t,z)$ and $\hat{A}^{+}(t,z)$. In
(\ref{sun}) we proceed to normalized time $\theta=t_1/\tau_{r}$ and then (\ref{oul})
can be presented like:
\begin{eqnarray}
\hat{O}(t)&=&i\gamma\int_{-\infty}^{\infty}\tilde{h}(\theta)\hat{n}_{0}(t-\theta\tau_{r})
d\theta\nonumber\\ &=&\int_{-\infty}^{\infty}\emph{G}(\theta)\hat{n}_{0}
(t\smash{-}\theta\tau_{r})d\theta,
\end{eqnarray}
where $\widetilde{h}(\theta)=\tau_{r}h(\theta\tau_r)$ and $\emph{G}(\theta)=i\gamma
\widetilde{h}(\theta)$. For the reduction of the expressions (\ref{oul}) to the
normally ordered form the following theorem can be used:\\
\textbf{Theorem:}\hspace{0.1in} {\em Bose-operator $e^{\hat{O}(t)}$ can be
represented in the normally ordered form this way:}
\begin{equation}
e^{\hat{O}(t)}=\hat{\mathbf{N}}\exp{\left\{\int_{-\infty}^{\infty}
\left[e^{G(\theta)}-1\right]\hat{n}_{0}(t-\theta\tau_{r})d\theta\right\}}.\label{teor}
\end{equation}
The operators in the integral expression may be understood as the $c$-numbers. In
\cite{Loudon} the similar theorem in the spectral representation was formulated and
demonstrated and we mention only that the similar demonstration can be done also in
this case. In fact in \cite{Loudon} the integration limits are not defined and the
theorem has not a obvious applicability. The demonstration of (\ref{teor}) does not
represent the central objective of this article so we formulate the theorem in the
time-representation only. The average value of the $e^{\hat{O}(t)}$ is given by the
formula
\begin{equation}\label{pp}
\langle e^{\hat{O}(t)}\rangle=\exp{\left\{\int_{-\infty}^{\infty}
\left[e^{G(\theta)}-1\right]\bar{n}_{0}(t-\theta\tau_{r})d\theta\right\}}.
\end{equation}
\subsection{THE AVERAGE VALUES OF ~$e^{\hat{O}(t)}$ AND ~$e^{\hat{O}^{+}(t)}$} In
most of the experimental situations $\gamma\ll1$ which allows one to decompose the
integral expression in the (\ref{pp}) and to limit decomposition at terms having
order $\gamma^{2}$. Using this approach we have:
\begin{eqnarray}\label{Tic}
\langle e^{\hat{O}(t)}\rangle=\exp
&{}&\Bigl[\int_{-\infty}^{\infty}\emph{G}(\theta)\bar{n}_{0}(t-\theta\tau_{r})d\theta\nonumber\\
&{}&+\frac{1}{2}\int_{-\infty}^{\infty}\emph{G}^{2}(\theta)\bar{n}_{0}(t-\theta\tau_{r})d\theta\Bigl],
\end{eqnarray}
where
$\bar{n}_{0}(t)\smash{=}\langle\hat{n}_{0}(t)\rangle\smash{=}|\alpha_{0}(t)|^{2}$. It
is convenient to enter in further analyse the envelope of a pulse $\rho(t)$, so that
$\alpha_0(t)\smash{=}\rho(t)\alpha_0$. If the initial pulse has the gaussian form
then $\rho(0)\smash{=}1$. For simplicity we denote:
\begin{eqnarray}
\psi(t)=\gamma\int_{-\infty}^{\infty}\widetilde{h}(\theta)\bar{n}_{0}(t-\theta\tau_{r})d\theta,\\
\mu(t)=\frac{1}{2}\gamma^{2}\!\!\int_{-\infty}^{\infty}\widetilde{h}^{2}(\theta)\bar{n}_{0}(t-\theta\tau_{r})d\theta.
\end{eqnarray}
{}From (\ref{Tic}) we have
\begin{equation}\label{sis}
\langle e^{\hat{O}(t)}\rangle=e^{i\psi(t)-\mu(t)},\quad
\langle e^{\hat{O}^{+}(t)}\rangle=e^{-i\psi(t)-\mu(t)}.
\end{equation}
The parameters $\psi(t)$ and $\mu(t)$ are connected with the self-action effect and
$\psi(t)$ represents nonlinear phase addition. Then
\begin{eqnarray}
\psi(t)=\psi_{0}\int_{0}^{\infty}\widetilde{h}(\theta)
\rho^{2}(t-\theta\tau_{r})d\theta\label{psi},\\
\mu(t)=\mu_{0}\int_{0}^{\infty}\widetilde{h}^{2}(\theta)\rho^{2}
(t-\theta\tau_{r})d\theta\label{mi},
\end{eqnarray}
where $\psi_{0}=2\gamma\alpha^{2}_{0}=2\gamma\bar{n}_{0}$ and
$\mu_{0}=\gamma^{2}\bar{n}_{0}=\gamma\psi_0/2$. A special interest is represented by
the estimation of the average values of the Bose-operator combinations at coherent
initial states (see (\ref{oul})). Taking into account (\ref{oul}) we have:
\begin{equation}
e^{\hat{O}(t_{1})}e^{\hat{O}(t_{2})}\smash{=}e^{\hat{O}(t_{2})}e^{\hat{O}
(t_{1})}\smash{=}e^{\hat{O}(t_{1})+\hat{O}(t_{2})}\smash{=}e^{\hat{Q}(t_{1},t_{2})}.
\end{equation}
In consequence we find
\begin{equation}
\hat{\mathbf{N}}\left[e^{\hat{O}(t_{1})}e^{\hat{O}(t_{2})}\right]=
\hat{\mathbf{N}}\left[e^{\hat{Q}(t_{1},t_{2})}\right],
\end{equation}
where:
\begin{eqnarray}
\hat{Q}(t_{1},t_{2})&=&\hat{O}(t_{1})+\hat{O}(t_{2})\nonumber\\
&=&\int_{-\infty}^{\infty}\!\!\widetilde{\emph{G}}
(t_{1},t_{2};\theta)\hat{n}_{0}(\theta\tau_{r})d\theta,\label{Qu}\\
\widetilde{\emph{G}}(t_{1},t_{2};\theta)&=&
i\gamma\widetilde{h}(t_{1}-\theta\tau_{r})+
i\gamma\widetilde{h}(t_{2}-\theta\tau_{r}).\label{Ka}
\end{eqnarray}
Using the theorem of normal ordering for (\ref{Qu}) we estimate the averages of
different combinations of Bose-operators:
\begin{eqnarray}
\langle e^{\hat{O}(t_{1})+\hat{O}(t_{2})}\rangle &=&
e^{i[\psi(t_{1})+\psi(t_{2})]-\mu(t_{1},t_{2})-\emph{K}(t_{1},t_{2})},\nonumber\\
\langle e^{\hat{O}^{+}(t_{1})+\hat{O}(t_{2})}\rangle &=&
e^{i[-\psi(t_{1})+\psi(t_{2})]-\mu(t_{1},t_{2})+\emph{K}(t_{1},t_{2})},\nonumber\\
\langle e^{\hat{O}(t_{1})+\hat{O}^{+}(t_{2})}\rangle &=&
e^{i[\psi(t_{1})-\psi(t_{2})]-\mu(t_{1},t_{2})+\emph{K}(t_{1},t_{2})},\label{tac}\\
\langle e^{\hat{O}^{+}(t_{1})+\hat{O}^{+}(t_{2})}\rangle &=&
e^{-i[\psi(t_{1})+\psi(t_{2})]-\mu(t_{1},t_{2})-\emph{K}(t_{1},t_{2})},\nonumber
\end{eqnarray}
where $\mu(t_{1},t_{2})\smash{=}\mu(t_{1})\smash{+}\mu(t_{2})$ and
$\emph{K}(t_{1},t_{2})$ represents the temporal correlator
\begin{equation}\label{fifa}
\emph{K}(t_{1},t_{2})=\mu_{0}\int_{-\infty}^{\infty}\widetilde{h}
(t_{1}\smash{-}\theta\tau_{r})\widetilde{h}(t_{2}\smash{-}\theta\tau_{r})
\rho^{2}(\theta\tau_{r})d\theta.
\end{equation}
In agreement with most of the experimental situations the approximation
$\tau_{p}\smash{\gg}\tau_{r}$ can be used. In this case in (\ref{psi}),(\ref{mi}) and
(\ref{fifa}) we can suppose that in time the envelope of a pulse slowly change itself
so it practically does not depend on the change of the integration variable. Therefore
it is possible to eliminate it from the under integral expression in the essential point
$\theta\tau_r=0$ in (\ref{psi},\ref{mi}) and $\theta\tau_r=t_1\smash{+}\tau/2$ in
(\ref{fifa}) which corresponds to the maximal value of under integral expression
$\widetilde{h}(\theta)=1$ (\ref{psi},\ref{mi}) and
$\widetilde{h}(t_{1}\smash{-}\theta\tau_{r})\widetilde{h}(t_{2}\smash{-}\theta\tau_{r})\smash{=}
\widetilde{h}(-\tau/2)\widetilde{h}(\tau/2)\smash{=}\widetilde{h}^2(\tau/2)$
(\ref{fifa}) consequently. Then
\begin{eqnarray}
\psi(t)&=&\psi_{0}\rho^{2}(t)\int_{0}^{\infty}\widetilde{h}(\theta)d\theta,\label{ps}\\
\mu(t)&=&\mu_{0}\rho^{2}(t)\int_{0}^{\infty}\widetilde{h}^{2}(\theta)d\theta,\label{miu}\\
\emph{K}(t_{1},t_{2})&=&\mu_{0}\rho^{2}(t_{1}+\tau/2)\frac{1}{\tau_{r}}
\int_{-\infty}^{\infty}\!\!
\widetilde{h}(\theta)\widetilde{h}(\theta+\tau)d\theta.\label{ku}
\end{eqnarray}
Here $\tau\smash{=}t_{2}\smash{-}t_{1}$. One should note that in our previous analyse
we did not choose the relaxation function of nonlinearity in a definite form. If the
nonlinearity is of a Kerr type, the relaxation function should be introduced as
\begin{equation}\label{brus}
H(t)=(1/\tau_r)e^{-t/\tau_r}\quad(t\ge0).
\end{equation}
Then $\widetilde{h}(\theta)\smash{=}e^{-|\theta|}$ and for the integrals in
(\ref{ps})-(\ref{ku}) we find
\begin{equation}
\int_{0}^{\infty}\widetilde{h}(\theta)d\theta=1,\qquad
\int_{0}^{\infty}\widetilde{h}^{2}(\theta)d\theta=\frac{1}{2},
\end{equation}
\begin{equation}\label{gas}
g(\tau)=\frac{1}{\tau_r}\int_{-\infty}^{\infty}\widetilde{h}(\theta)\widetilde
{h}(\theta\smash{+}\tau)d\theta=\frac{1}{\tau_r}\Bigl(1\smash{+}\frac{|\tau|}{\tau_{r}}
\Bigl)\widetilde{h}\Bigl(\frac{\tau}{\tau_{r}}\Bigl).
\end{equation}
\section{THE CORRELATION FUNCTION OF QUADRATURES}
As stated earlier (see (\ref{nor1})) in self-action process the photon statistics
remains unchanged. Therefore, we are interested in analysing the quadrature
components which are defined as:
\begin{eqnarray}
\hat{X}(t,z)&=&\left[\hat{A}(t,z)+\hat{A}^{+}(t,z)\right]/2,\\
\hat{Y}(t,z)&=&\left[\hat{A}(t,z)-\hat{A}^{+}(t,z)\right]/2i.
\end{eqnarray}
The averages of the operators $\hat{A}(t,z)$ and $\hat{A}^{+}(t,z)$ at initial
coherent state of a pulse are
\begin{eqnarray}
\langle\hat{A}(t,z)\rangle&=&\alpha_{0}(t)\langle e^{\hat{O}(t)}\rangle,\\
\langle\hat{A}^{+}(t,z)\rangle&=&\alpha^{*}_{0}(t)\langle e^{\hat{O}^{+}(t)}\rangle.
\end{eqnarray}
Taking into account (\ref{sis}) and that
$\alpha_{0}(t)\smash{=}|\alpha_{0}(t)|e^{i\varphi(t)}$, for average values of
quadratures we obtain:
\begin{eqnarray}
\langle\hat{X}(t,z)\rangle&=&|\alpha_{0}(t)|e^{-\mu(t)}\cos{\Phi(t)},\label{xip}\\
\langle\hat{Y}(t,z)\rangle&=&|\alpha_{0}(t)|e^{-\mu (t)}\sin{\Phi(t)}.\label{xix}
\end{eqnarray}
where $\Phi(t)=\psi(t)\smash{+}\varphi(t)$. Exponential term in (\ref{xip}) and
(\ref{xix}) is caused by quantum effects - in the classical theory it is not present.
{}From (\ref{xip})-(\ref{xix}) it is concluded that the changes of quadratures in
time are connected with changes in pulse's envelope. We introduce correlation
functions of quadrature components as
\begin{eqnarray}
R_{X}(t,t\smash{+}\tau)\!\!=\!\!\frac{1}{2}\!\!&\Bigl[&\!\!\!
\langle\hat{X}(t,z)\hat{X}(t\smash{+}\tau)\rangle\smash{+}
\langle\hat{X}(t\smash{+}\tau,z)\hat{X}(t,z)\rangle\nonumber \\
&-&2\langle\hat{X}(t,z)\rangle\langle\hat{X}(t\smash{+}\tau,z)\rangle\Bigl],\\
R_{Y}(t,t\smash{+}\tau)\!\!=\!\!\frac{1}{2}\!\!&\Bigl[&\!\!\!
\langle\hat{Y}(t,z)\hat{Y}(t\smash{+}\tau)\rangle\smash{+}\langle
\hat{Y}(t\smash{+}\tau,z)\hat{Y}(t,z)\rangle\nonumber \\
&-&2\langle\hat{Y}(t,z)\rangle\langle\hat{Y}(t\smash{+}\tau,z) \rangle\Bigl].
\end{eqnarray}
To analyse the correlation functions of quadrature components, it is necessary to
evaluate the correlators
$\xi_{X}(t_{1},t_{2})\smash{=}\langle\hat{X}(t_{1})\hat{X}(t_{2})\rangle$ and
$\xi_{Y}(t_{1},t_{2})\smash{=}\langle\hat{Y}(t_{1})\hat{Y}(t_{2})\rangle$. Using
permutation relations (\ref{Tab}) and (\ref{tac}), we obtain
\begin{eqnarray}
\xi_{X}(t_{1},t_{2})&=&\frac{1}{4}\delta(t_{2}-t_{1})+\frac{1}{2}|\alpha_{0}(t_{1})|
|\alpha_{0}(t_{2})|e^{-\mu(t_{1},t_{2})}\nonumber \\
&{}&\times\Bigl[e^{-\Lambda(t_{1},t_{2})}
\cos[\Phi(t_{1})\smash{+}\Phi(t_{2})\smash{+}
\gamma\widetilde{h}(t_{2}\smash{-}t_{1})]\nonumber\\
&{}&\,\,+e^{\Lambda(t_{1},t_{2})}\cos[\Phi(t_{1})\smash{-}\Phi(t_{2})]\Bigl],\\
\xi_{Y}(t_{1},t_{2})&=&\frac{1}{4}\delta(t_{2}-t_{1})-\frac{1}{2}|\alpha_{0}(t_{1})|
|\alpha_{0}(t_{2})|e^{-\mu(t_{1},t_{2})}\nonumber\\
&{}&\times\Bigl[e^{-\Lambda(t_{1},t_{2})}\cos[\Phi(t_{1})\smash{+}
\Phi(t_{2})\smash{+}\gamma\widetilde{h}(t_{2}\smash{-}t_{1})]\nonumber\\
&{}&\,\,-e^{\Lambda(t_{1},t_{2})}\cos[\Phi(t_{1})\smash{-}\Phi(t_{2})]\Bigl],
\end{eqnarray}
where $\Lambda(t_{1},t_{2})\smash{=}\mu(t_1,t_2)\widetilde{h}(t_2\smash{-}t_1)$ (see
(\ref{Ka})). As a result for the correlation functions of quadratures we have:
\begin{eqnarray}
R_{X}(t,t&+&\tau)=\frac{1}{4}\Bigl[\delta(\tau)\nonumber\\
&-&\!\psi_{0}\rho(t)\rho(t\smash{+}\tau)h(\tau)\sin[\Phi(t)+\Phi(t\smash{+}\tau)]\nonumber\\
&+&\!\psi^{2}_{0}\rho(t)\rho(t\smash{+}\tau)g(t,\tau)\sin{\Phi(t)}\sin{\Phi(t\smash{+}\tau)}\Bigl],\label{rer}\\
R_{Y}(t,t&+&\tau)=\frac{1}{4}\Bigl[\delta(\tau)\nonumber\\
&+&\!\psi_{0}\rho(t)\rho(t\smash{+}\tau)h(\tau)\sin[\Phi(t)+\Phi(t\smash{+}\tau)]\nonumber\\
&+&\!\psi^{2}_{0}\rho(t)\rho(t\smash{+}\tau)g(t,\tau)\cos{\Phi(t)}\cos{\Phi(t\smash{+}\tau)}\Bigl],\label{yer}
\end{eqnarray}
where
\begin{equation}\label{geps}
g(t,\tau)\!=\!\rho^{2}(t+\tau/2)g(\tau).
\end{equation}
To obtain (\ref{rer}) and (\ref{yer}) the $\gamma\ll1$ and $\tau_{r}\ll\tau_{p}$
approximations have been used.
\section{THE SPECTRUM OF QUANTUM FLUCTUATIONS OF QUADRATURE COMPONENTS} Spectral
densities of fluctuations of the quadratures are defined by the following
expressions:
\begin{eqnarray}
S_{X}(\omega,t)&=&\int_{-\infty}^{\infty}R_{X}(t,t+\tau)e^{i\omega\tau}d\tau,\\
S_{Y}(\omega,t)&=&\int_{-\infty}^{\infty}R_{Y}(t,t+\tau)e^{i\omega\tau}d\tau.
\end{eqnarray}
Taking into account the weak change of the envelope during the relaxation time
 one obtains:
\begin{eqnarray}
S_{X}(\omega,t)&=&\frac{1}{4}\Bigl[1-\psi_{0}\rho^{2}(t)\sin{2\Phi(t)}
\int_{-\infty}^{\infty}h(\tau)e^{i\omega\tau}d\tau\nonumber\\
&{}&+\psi^{2}_{0}\rho^{4}(t)\sin^{2}{\Phi(t)}\int_{-\infty}^{\infty}
g(\tau)e^{i\omega\tau}d\tau\Bigl],\label{soi}\\
S_{Y}(\omega,t)&=&\frac{1}{4}\Bigl[1+\psi_ {0}\rho^{2}(t)\sin{2\Phi(t)}
\int_{-\infty}^{\infty}h(\tau)e^{i\omega\tau}d\tau\nonumber\\
&{}&+\psi^{2}_{0}\rho^{4}(t)\cos^{2}{\Phi(t)}\int_{-\infty}^{\infty}
g(\tau)e^{i\omega\tau}d\tau\Bigl].\label{pio}
\end{eqnarray}
The estimation of integrals in (\ref{soi}),(\ref{pio}) gives us
\begin{eqnarray}
\int_{-\infty}^{\infty}\!\!h(\tau)e^{i\omega\tau}d\tau
&=&\frac{2}{1+(\omega\tau_{r})^{2}}=2L(\Omega),\label{cipi1}\\
\int_{-\infty}^{\infty}\!\!g(\tau)e^{i\omega\tau}d\tau
&=&\frac{4}{[1+(\omega\tau_{r})^{2}]^{2}}=4L^{2}(\Omega),\label{cipi2}
\end{eqnarray}
where $\Omega=\omega\tau_{r}$. Hence
\begin{eqnarray}
S_{X}(\Omega,t)=\frac{1}{4}\Bigl[1&-&2\psi(t)L(\Omega)\sin{2\Phi(t)}\nonumber\\
&{}&+4\psi^{2}(t)L^{2}(\Omega)\sin^{2}{\Phi(t)}\Bigl],\label{sio}\\
S_{Y}(\Omega,t)=\frac{1}{4}\Bigl[1&+&2\psi(t)L(\Omega)\sin{2\Phi(t)}\nonumber\\
&{}&+4\psi^{2}(t)L^{2}(\Omega)\cos^{2}{\Phi(t)}\Bigl],\label{ert}
\end{eqnarray}
where $\psi(t)\smash{=}2\gamma|\alpha_{0}(t)|^{2}$. From (\ref{sio}) and (\ref{ert})
follows that the choice of the phase $\Phi(t)$ determines the level of quantum
fluctuations lower and higher than the shot-noise level
$S_{X}(\omega)\smash{=}S_{Y}(\omega)\smash{=}1/4$, corresponding to the coherent
state of the initial pulse. In conformity with the Heisenberg relation the behaviour
of the spectrum of the $X$-quadrature appears to be moved with
 a phase $\pi/2$ in comparison
with $Y$-quadrature. In case of an optimal phase of the initial pulse
\begin{equation}\label{phase}
\varphi_{0}(t)=\frac{1}{2}\arctan{\left[\frac{1}{\psi(t)L(\Omega_{0})}\right]}-
\psi(t)
\end{equation}
chosen for the frequency $\Omega_{0}\smash{=}\omega_{0}\tau_{r}$, spectral densities
\begin{eqnarray}
S_{X}(\Omega_{0},t)\!&=&\!\!\frac{1}{4}\Bigl[\sqrt{1\!+
\!\psi^{2}(t)L^{2}(\Omega_{0})}\!-\!\psi(t)L(\Omega_{0})\Bigl]^{2},\label{So}\\
S_{Y}(\Omega_{0},t)\!&=&\!\!\frac{1}{4}\Bigl[\sqrt{1\!+
\!\psi^{2}(t)L^{2}(\Omega_{0})}\!+\!\psi(t)L(\Omega_{0})\Bigl]^{2}.\label{Soe}
\end{eqnarray}
Eqs.(\ref{So}),(\ref{Soe}) indicate that when the nonlinear phase addition $\psi(t)$
increases $S_{X}(\Omega_{0},t)$ monotonously decreases and $S_{Y}(\Omega_{0},t)$
monotonously increases. At any frequency $\Omega$ we have
\begin{eqnarray}
&S&\!_{X}(\Omega,t)=S_{X}(\Omega_{0},t)+\frac{1}{2}\psi(t)[L(\Omega)-L(\Omega_{0})]\nonumber\\
&\times&\!\Bigl\{[L(\Omega)\smash{+}L(\Omega_{0})]\psi(t)\smash{-}[1\smash{+}(L(\Omega)\smash{+}
L(\Omega_{0}))L(\Omega_{0})\psi^{2}(t)]\nonumber\\
&{}&\times[1\smash{+}\psi^{2}(t)L^{2}(\Omega)]^{-1/2}\Bigl\},\label{sist}\\
&S&\!_{Y}(\Omega,t)=S_{Y}(\Omega_{0},t)+\frac{1}{2}\psi(t)[L(\Omega)-L(\Omega_{0})]\nonumber\\
&\times&\!\Bigl\{[L(\Omega)\smash{+}L(\Omega_{0})]\psi(t)\smash{+}[1\smash{+}(L(\Omega)\smash{+}
L(\Omega_{0}))L(\Omega_{0})\psi^{2}(t)]\nonumber\\
&{}&\times[1\smash{+}\psi^{2}(t)L^{2}(\Omega)]^{-1/2}\Bigl\},\label{sios}
\end{eqnarray}

\begin{figure}[t]
   \begin{center}
       \leavevmode
       \epsfxsize=.48\textwidth
       \epsfysize=.4\textwidth
       \epsffile{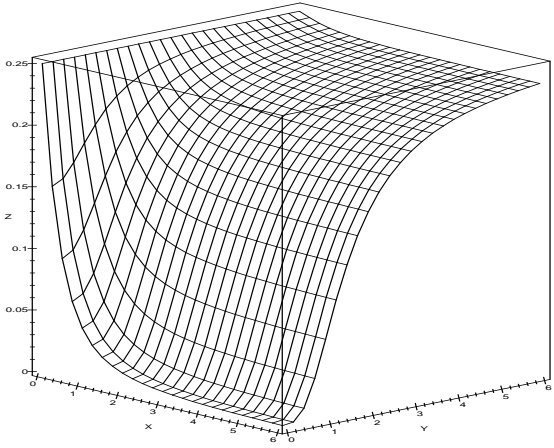}
   \end{center}
   \caption{Dispersion [Z] of the quadrature component of a pulse at time $t\smash{=}0$
as a function of the maximum nonlinear phase $\psi(0)\smash{=}2\gamma|\alpha^2(0)|$
[X] and the reduced frequency $\Omega\smash{=}\omega\tau_r$ [Y], at values of
the phase of the pulse which are optimal for $\Omega\smash{=}0$.\label{su1}}
\end{figure}

\begin{figure}[t]
    \begin{center}
        \leavevmode
        \epsfxsize=.48\textwidth
        \epsfysize=.4\textwidth
        \epsffile{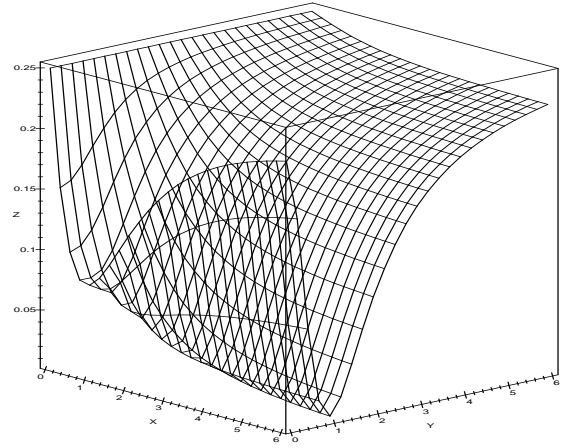}
    \end{center}
\caption{Dispersion [Z] of the quadrature component of a pulse at time $t\smash{=}0$
as a function of the maximum nonlinear phase $\psi(0)\smash{=}2\gamma|\alpha^2(0)|$
[X] and the reduced frequency $\Omega\smash{=}\omega\tau_r$ [Y], at values of
the phase of the pulse which are optimal for $\Omega\smash{=}1$.\label{su2}}
\end{figure}

The spectra of $X$-quadrature component, calculated by the formula (\ref{sist}), at
$t\smash{=}0$ ($\psi(0)\smash{=}\psi_{0}$) for the cases
$\Omega_{0}\smash{=}0~(\omega_{0}\smash{=}0)$,
$\Omega_{0}\smash{=}1~(\omega_{0}\smash{=}\tau^{-1}_{r})$ are presented in Figs.\
\ref{su1},\ref{su2} respectively. On Fig.\ \ref{su1} one can see that for
$\omega_{0}\smash{=}0$ spectral density of $X$-quadrature component is minimal on
frequency $\omega\smash{=}0$ for any values of phase $\psi_{0}$. For
$\omega_{0}\smash{\neq}0$ (Fig.\ \ref{su2}) and phases $\psi_{0}\smash{>}1$ the
minimum of the fluctuation spectrum of $X$-quadrature component lies at frequencies
$\Omega\smash{=}1~(\omega\smash{=}\tau^{-1}_{r})$, and for  $\psi_{0}\smash{<}1$ the
minimum lies near $\Omega\smash{\approx}0~(\omega\smash{\approx}0)$.
\section{THE WIDTH OF THE SPECTRUM OF SQUEEZED QUADRATURE}
\begin{figure}[t]
     \begin{center}
       \leavevmode
       \epsfxsize=.48\textwidth
       \epsfysize=.37\textwidth
       \epsffile{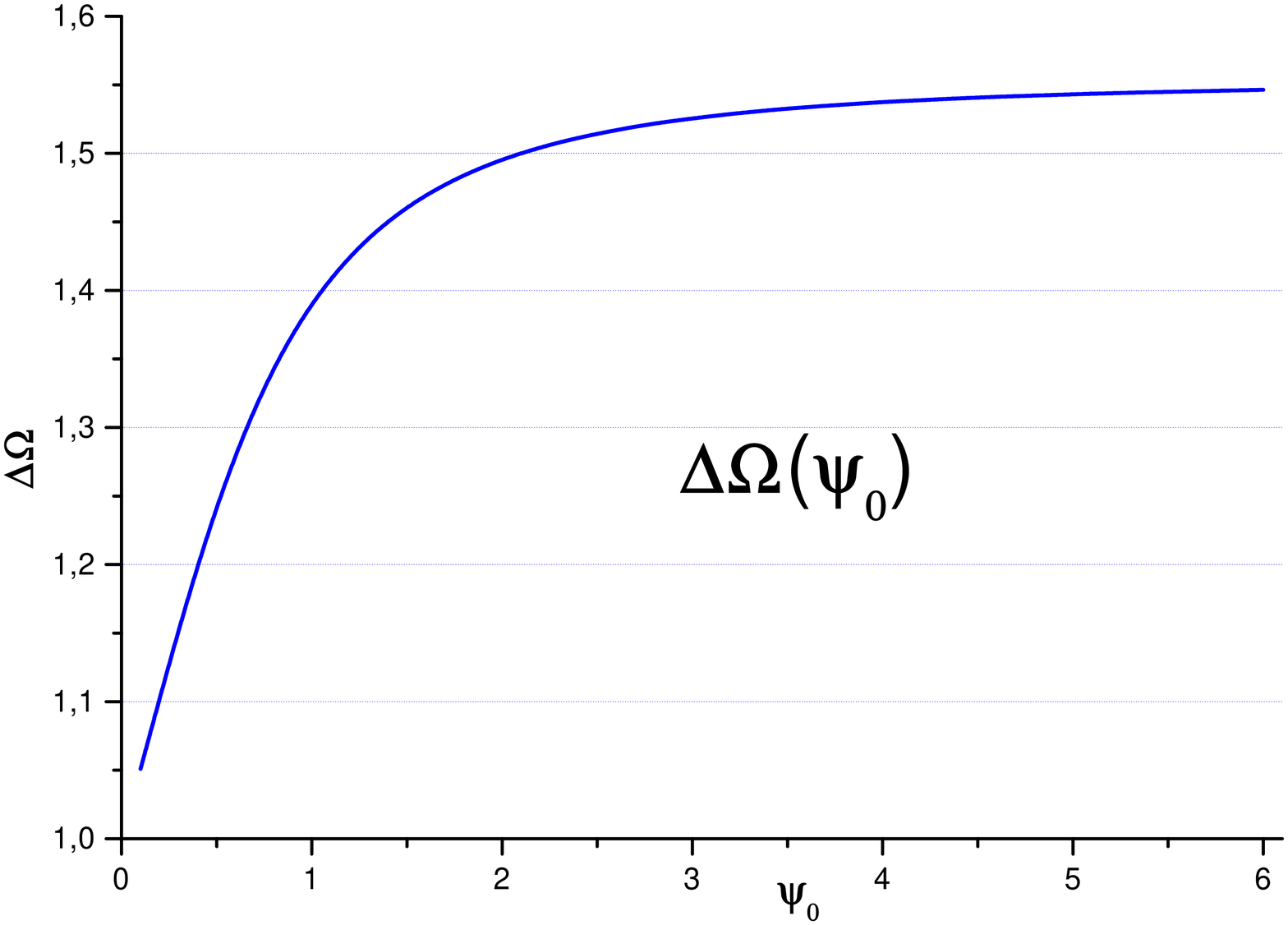}
     \end{center}
     \caption{Spectral band $\Delta\Omega\smash{=}2\tau_{r}\Delta\omega$
of the quadrature of a pulse with suppressed quantum fluctuations as a function of
the maximum nonlinear phase $\psi_{0}$.\label{su3}}
\end{figure}

{}From Fig.\ \ref{su1} one can conclude that the frequency where spectral density of
"$•$-quadrature fluctuations" is lower than the shot-noise level, depends on the
nonlinear phase addition $\psi_{0}$. Width of the spectrum below the shot-noise level
$\Delta\Omega\smash{=}\tau_r\Delta\omega$ should be defined from
\begin{equation}\label{glub}
S_{X}(\Delta\Omega,t)=\frac{1}{2}\Bigl[\frac{1}{4}+S_{X}(\Omega_{0},t)\Bigl].
\end{equation}
Accounting (\ref{So},\ref{sist}) for $\Omega_{0}\smash{=}0$ from (\ref{glub}) we
have:
\begin{eqnarray}
2\psi(t)\Bigl[\psi(t)&-&\sqrt{1+\psi^{2}(t)}~\Bigl]
L^{2}(\Delta\Omega)+2L(\Delta\Omega)\nonumber\\
&+&\psi(t)\sqrt{1+\psi^{2}(t)}-\psi^{2}(t)-1=0.\label{rtc}
\end{eqnarray}
Eq.(\ref{rtc}) in $L(\Delta\Omega)$ (see (\ref{cipi1},\ref{cipi2})) has two solution
of which only one is real. Solving the (\ref{rtc}) for $\Delta\Omega$ we get:
\begin{eqnarray}\label{ela}
\Delta\Omega&=&\Biggl[\frac{2\psi(t)\Bigl[\psi(t)-\sqrt{1+\psi^{2}(t)}~\Bigl]}
{-1+\sqrt{1-2\psi(t)\Bigl[\psi(t)-\sqrt{1+\psi^{2}(t)}~\Bigl]}}\nonumber\\
&{}&\times\frac{1}{\sqrt{\Bigl[\psi(t)\sqrt{1+\psi^{2}(t)}-\psi^{2}(t)-1\Bigl]}}
-1\Biggl]^{1/2}.
\end{eqnarray}
{}From (\ref{ela}) it follows that the change of $\Delta\Omega$ is connected with
changes in pulse's envelope. The frequency band in which the spectral density of the
quadrature fluctuations is lower than the shot-noise level depends on the nonlinear
phase shift $\psi(t)$. The corresponding dependence at $t\smash{=}0$ for
$\omega\smash{=}0$ is displayed in Fig.\ \ref{su3}. It may be noted that at
$\psi_{0}\smash{\gg}1$ width of the spectrum below the shot noise level is one and a
half width of the spectral response of nonlinearity.
\section{PHOTON NUMBER SPECTRAL``DENSITY'' OF PULSES WITH SELF-PHASE MODULATION}
The spectral photon number operator is defined by
\begin{equation}
\hat{n}(\omega,z)=\hat{a}^+(\omega,z)\hat{a}(\omega,z),\label{auto}
\end{equation}
where:
\begin{eqnarray}
\hat{a}(\omega,z)&=&\frac{1}{\sqrt{2\pi}\,\tau_p}\int_{-\infty}^{\infty}
\hat{A}(t,z)e^{i\omega t}dt,\label{alpi1}\\
\hat{a}^+(\omega,z)&=&\frac{1}{\sqrt{2\pi}\,\tau_p}\int_{-\infty}^{\infty}
\hat{A}^+(t,z)e^{-i\omega t}dt.\label{alpi2}
\end{eqnarray}
Taking into account (\ref{alpi1},\ref{alpi2}) and (\ref{sta1},\ref{sta2}) for photon
number spectral ``density'' (\ref{auto}) we find:
\begin{eqnarray}\label{gori}
\bar{n}(\omega,z)&=&\frac{1}{2\pi\tau^2_p}\int_{-\infty}^{\infty}\int_{-\infty}
^{\infty}\langle\hat{A}^+_0(t_1)\,e^{\hat{O}^+(t_1)
+\hat{O}(t_2)}\hat{A}_0(t_2)\rangle\nonumber\\ &{}&\times
e^{i\omega(t_2-t_1)}dt_1dt_2\cong\frac{\bar{n}_0}
{2\pi}\cdot\frac{1}{\tau^2_p}\cdot|\mathbf{I}|^2,
\end{eqnarray}
where
\begin{equation}\label{lom}
{\mathbf{I}}=\int_{-\infty}^{\infty}\rho(t)\,
\exp{\left\{i\left[\psi_0\rho^2(t)+\varphi(t)+\omega t\right]\right\}}dt.
\end{equation}
The last expression in (\ref{gori}) is written without the account of $\mu(t_1,t_2)$
and $\emph{K}(t_1,t_2)$ (see (\ref{tac})). If the initial pulse has gaussian form
$\rho(t)\smash{=}\exp{\{-t^2/2\tau^2_p\}}$ then, using paraxial approximation
($\rho^2(t)\cong1\smash{-}t^2/\tau^2_p$) \cite{Popescu2} in (\ref{lom}) for spectral
density (\ref{gori}) we find
\begin{equation}\label{su25}
\bar{n}_{class}(\Omega,z)=\frac{\bar{n}_0}{\sqrt{1+4\psi^2_0}}
\exp{\left[-\frac{\Omega^2}{1+4\psi^2_0}\right]},
\end{equation}
where $\Omega\smash{=}\omega\tau_p$. From (\ref{su25}) it follows that the spectral
density of a pulse with self-phase modulation (SPM-USP) decreases when nonlinear
phase addition increases. To calculate (\ref{su25}) in (\ref{gori}) the terms
$\mu(t_1,t_2)$ and $K(t_1,t_2)$ have not been taken into account. In consequence, the
spectral density (\ref{su25}) does not depend on relaxation time. If we take the
relaxation function of nonlinearity as \cite{Popescu2}
\begin{equation}\label{gicu}
\emph{H}(\theta)=\frac{1}{\tau_{r}}\exp{\left\{-\frac{\theta^2}{2\tau^2_r}\right\}}
\end{equation}
then
$\emph{K}(t,t\smash{+}\tau)=\mu_0\rho^2(t\smash{+}\tau/2)\exp{\{-\tau^2/4\tau^2_r\}}$.
Using the expression $\emph{K}(t,t+\tau)$ in (\ref{gori}) in paraxial approximation
we have
\begin{eqnarray}\label{optics}
\bar{n}(\Omega,z)\!&=&\!\frac{\bar{n}_0}{\left[\left(1+\gamma\psi_0\nu^2/4\right)^2
+4\psi^2_0\right]^{1/2}}\nonumber\\
&{}&\times\exp{\left[-\frac{\Omega^2\left(1+\gamma\psi_0\nu^2/4\right)}
{\left(1+\gamma\psi_0\nu^2/4\right)^2+4\psi^2_0}\right]},
\end{eqnarray}
where $\nu\smash{=}\tau_p/\tau_r$. From (\ref{optics}) one can conclude that the
spectral density depends on the nonlinear phase addition and on
the relation between the
pulse duration and the relaxation time of the nonlinearity.
\section{THE CORRELATION FUNCTION OF SPECTRAL COMPONENTS OF SPM-USP{\scriptsize s}}
We introduce the correlation function of different spectral components in the
following symmetric form:
\begin{eqnarray}\label{pol}
R(\omega_1,\omega_2,z)\!\!&=&\!\!\frac{1}{2}\Bigl[\langle\hat{n}(\omega_1,z)\hat{n}(\omega_2,
z)\rangle\!+\!\langle\hat{n}(\omega_2,z)\hat{n}(\omega_1,z)\rangle\!\nonumber\\
&{}&-2\langle\hat{n}(\omega_1,z)\rangle\langle\hat{n}(\omega_2,z)\rangle\Bigl].
\end{eqnarray}
Leaving out the preliminary accounts for (\ref{pol}) we obtain
\begin{equation}\label{ret}
R(\omega_1,\omega_2,z)=I_1\delta(\omega_2\smash{-}\omega_1)-\frac{1}{2}\psi_0Im\{I^*_2I_3+
I_2I^*_3\},
\end{equation}
\begin{mathletters}
\begin{eqnarray}
I_1=\int_{-\infty}^{\infty}\int_{-\infty}^{\infty}&{}&
\widetilde{\rho}(t_1,t_2)\exp{\{i[\psi^2(t_1)-\psi^2(t_2)]\}}\nonumber\\
&{}&\times\exp{i[\omega_1t_1-\omega_2t_2]}dt_1dt_2,\\ \label{venera}
I_2=\int_{-\infty}^{\infty}\int_{-\infty}^{\infty}&{}&
\widetilde{\rho}(t_1,t_2)\exp{\{i[\psi^2(t_1)+\psi^2(t_2)]\}}\nonumber\\
&{}&\times\exp{\{i[\omega_1t_1+\omega_2t_2]\}}dt_1dt_2,\\ \label{sirius}
I_3=\int_{-\infty}^{\infty}\int_{-\infty}^{\infty}&{}&
\widetilde{q}(t_1,t_2)\exp{\{i[\psi^2(t_1)+\psi^2(t_2)]\}}\nonumber\\
&{}&\times\exp{\{i[\omega_1t_1+\omega_2t_2]\}}dt_1dt_2.\label{primacov}
\end{eqnarray}
\end{mathletters}
$\widetilde{\rho}(t_1,t_2)\smash{=}\rho(t_1)\rho(t_2)$,
$\widetilde{q}(t_1,t_2)\smash{=}\widetilde\rho(t_1,t_2)\widetilde{h}(t_2\smash{-}t_1)$.
If the initial pulse has gaussian form and the relaxation function has the form
(\ref{gicu}) then in paraxial approximation \cite{Popescu2} for correlation function
(\ref{ret}) we find
\begin{equation}\label{greu}
R(\Omega_1,\Omega_2,z)\!=\!I_1\delta(\Omega_1\smash{-}\Omega_2)
\smash{-}\frac{\psi_0}{2\widetilde{\alpha}
\sqrt{\widetilde{\beta}}}Im\{\Gamma(\Omega_1,\Omega_2)\},
\end{equation}
\begin{equation}\label{i1}
I_1=[n_{class}(\Omega_1)n_{class}(\Omega_2)]^{1/2}
\exp{\Bigl\{i\psi_0\frac{\Omega^2_1-\Omega^2_2}{1+4\psi^2_0}\Bigl\}}.
\end{equation}
In (\ref{greu}) are entered the following designations:
\begin{equation}
\Omega=\omega\tau_p,\qquad\widetilde{\alpha}=\left[1+4\psi^2_0\right]^{1/2},
\end{equation}
\begin{equation}
\widetilde{\beta}=\left[1+2\nu^2-4\psi^2_0)^2+16(1+\nu^2)^2\psi^2_0\right]^{1/2},
\end{equation}
\begin{equation}\label{bobo}
{\large{\Gamma}}(\Omega_1,\Omega_2)=\exp{\{G+iS\}}+\exp{\{E+iF\}},
\end{equation}
\begin{equation}\label{sirius1}
Im\Gamma(\Omega_1,\Omega_2)=e^{G}\sin{S}+e^{E}\sin{F},
\end{equation}
\begin{mathletters}
\begin{eqnarray}\label{tica1}
G&=&-\frac{\Omega^2_1+\Omega^2_2}{2\widetilde{\alpha}}\cos{\epsilon}-\frac{\Omega^2_
2}{2\varrho}\cos{\Sigma}-\frac{\varrho\,\Omega^2_1}{2\widetilde
{\beta}}\cos{(\Sigma-\xi)}\nonumber\\ &{}&-\frac{\Omega_1\Omega_2}{\widetilde
{\beta}}\nu^2\cos{\xi}-\frac{\Omega^2_2\nu^4}{2\varrho\widetilde{\beta}}\cos
{(\Sigma+\xi)},\\ S&=&\frac{\Omega^2_1+\Omega^2_2}{2\widetilde{\alpha}}\sin{\epsilon}
+\frac{\Omega^2_2}{2\varrho}\sin{\Sigma}-\frac{\varrho\,\Omega^2_1}{2\widetilde{\beta}}
\sin{(\Sigma\smash{-}\xi)}\nonumber\\
&{}&+\frac{\Omega_1\Omega_2}{\widetilde{\beta}}\nu^2\sin{\xi}\smash{+}
\frac{\Omega^2_2\nu^4}{2\varrho\widetilde{\beta}}\sin{(\Sigma\smash{+}\xi)}
\smash{-}(\epsilon\smash{+}\frac{\xi}{2}),\\
E&=&-\frac{\Omega^2_1+\Omega^2_2}{2\widetilde{\alpha}}\cos{\epsilon}-
\frac{\Omega^2_1}{2\varrho}\cos{\Sigma}-\frac{\varrho\,\Omega^2_2}{2\widetilde{\beta}}
\cos{(\Sigma\smash{-}\xi)}\nonumber\\
&{}&-\frac{\Omega_1\Omega_2}{\widetilde{\beta}}\nu^2\cos{\xi}\smash{-}
\frac{\Omega^2_1\nu^4}{2\varrho\widetilde{\beta}}\cos{(\Sigma\smash{+}\xi)},\\
F&=&\frac{\Omega^2_1+\Omega^2_2}{2\widetilde{\alpha}}\sin{\epsilon}+
\frac{\Omega^2_1}{2\varrho}\sin{\Sigma}-\frac{\varrho\,\Omega^2_2}{2\widetilde{\beta}}
\sin{(\Sigma\smash{-}\xi)}\nonumber\\
&{}&+\frac{\Omega_1\Omega_2}{\widetilde{\beta}}\nu^2\sin{\xi}+
\frac{\Omega^2_1\nu^4}{2\varrho\widetilde{\beta}}\sin{(\Sigma\smash{+}\xi)}
-(\epsilon\smash{+}\frac{\xi}{2}),
\end{eqnarray}
\end{mathletters}
\begin{eqnarray}\label{tica4}
\varrho&=&\left[(1+\nu^2)^2+4\psi^2_0\right]^{1/2},\\
\epsilon&=&-\arctan{\left\{2\psi_0\right\}},\\
\xi&=&\arctan{\left\{\frac{4(1+\nu^2)\psi_0}{1+2\nu^2-4\psi^2_0}\right\}},\\
\Sigma&=&\arctan{\left\{\frac{2\psi_0}{1+\nu^2}\right\}}.
\end{eqnarray}
We define the spectral correlation function of the photons with frequency $\Omega$ in
the spectral band $\Delta\Omega$
\begin{eqnarray}\label{cip}
\widetilde{R}_{\Delta\Omega}(\Omega,z)&=&\int_{\Omega-
\normalsize{\Delta\Omega/2}}^{\Omega+\normalsize{\Delta\Omega/2}}
\int_{\Omega-\normalsize{\Delta\Omega/2}}
^{\Omega+\normalsize{\Delta\Omega/2}}R(\Omega_1,\Omega_2,z)
d\Omega_1d\Omega_2\nonumber\\ &{}& -\bar{n}_{0}.
\end{eqnarray}
As a consequence, the conclusion which one can make is that for
$\widetilde{R}(\Omega,z)\smash{<}0$ take place the photon antibunching and for
$\widetilde{R}(\Omega,z)\smash{>}0$ the photon bunching.

\begin{figure}[t]
    \begin{center}
        \leavevmode
         \epsfxsize=.48\textwidth
         \epsfysize=.37\textwidth
        \epsffile{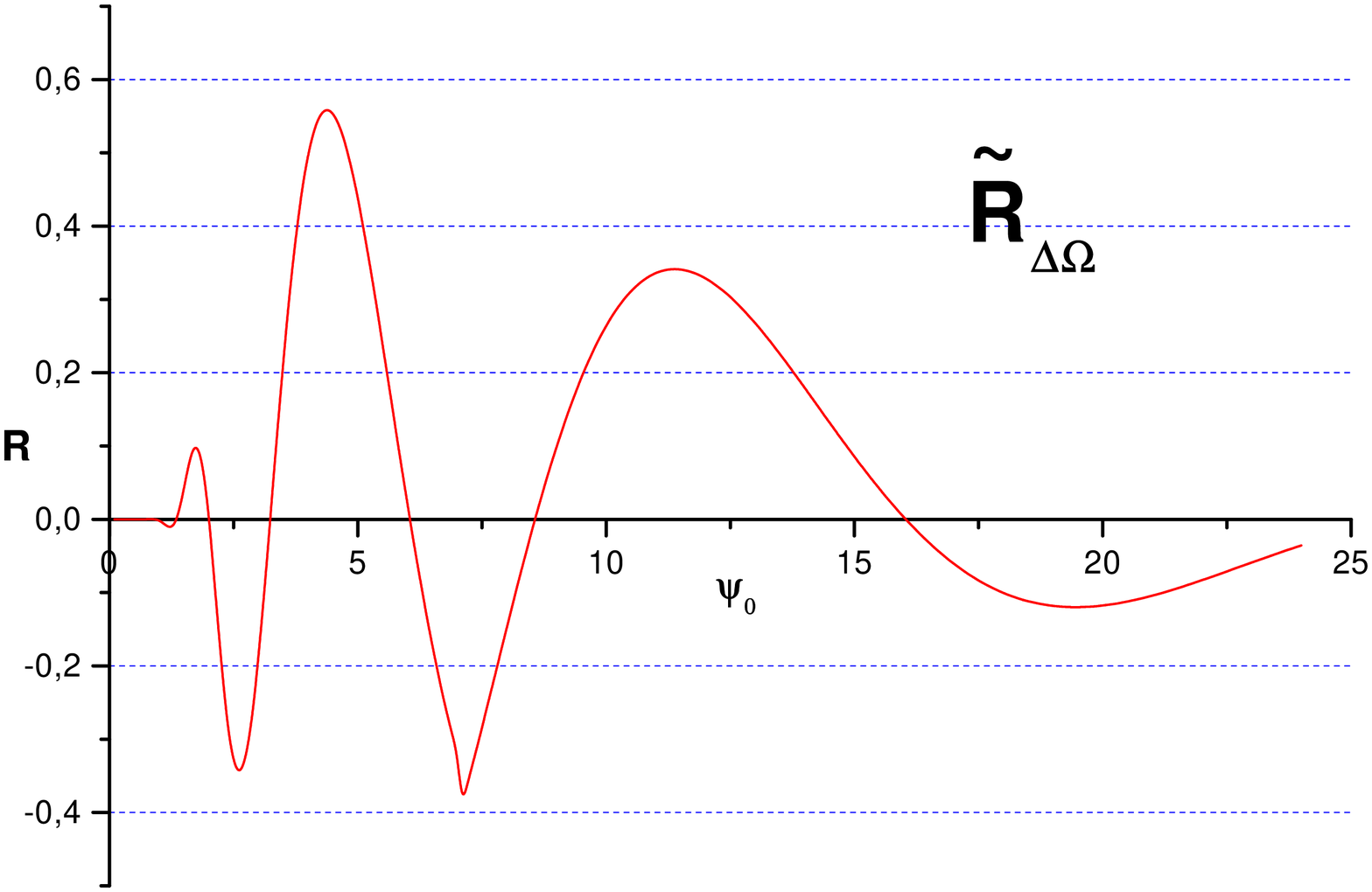}
     \end{center}
\caption{Spectral correlation function $\widetilde{R}_{\Delta\Omega}(\Omega,z)$ [R]
as a function of the maximum nonlinear phase $\psi_0$ measured at the frequency
$\Omega\smash{=}0.04$ in spectral band
$\Delta\Omega\smash{=}2.5\smash{\cdot}10^{-3}$.\label{su4}}
\end{figure}

\begin{figure}[t]
    \begin{center}
        \leavevmode
         \epsfxsize=.48\textwidth
         \epsfysize=.37\textwidth
        \epsffile{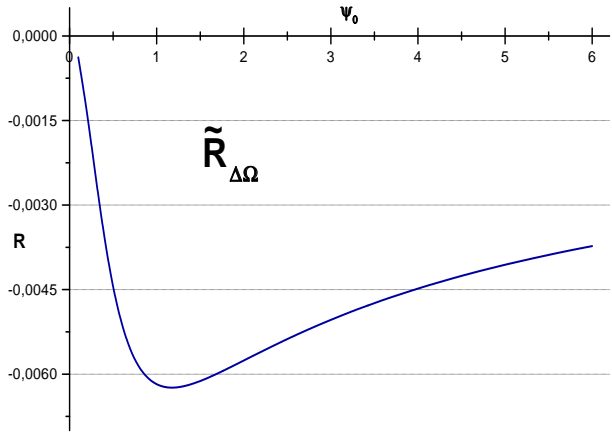}
     \end{center}
\caption{Spectral correlation function $\widetilde{R}_{\Delta\Omega}(\Omega,z)$ [R]
as a function of the maximum nonlinear phase $\psi_0$ measured at the frequency
$\Omega=0$ in spectral band $\Delta\Omega=0.75$.\label{su5}}
\end{figure}

The graphic dependence of the spectral correlation function (\ref{cip}) on $\psi_0$
at $\Omega\smash{=}0.04$, $\Delta\Omega\smash{=}2.5\smash{\cdot}10^{-3}$ and
$\nu\smash{=}10$ is displayed in Fig.\ref{su4}, whence it follows that the photon
bunching or antibunching can take place, and for phases $\psi_0\smash{>}1$ it becomes
significant.
At frequency $\Omega\smash{=}0$, the correlation function (see (\ref{cip})) has the
following simplified form:
\begin{equation}\label{deli}
\widetilde{R}_{\Delta\Omega}(0,z)=\frac{\psi_0}{2\widetilde{\alpha}
\sqrt{\widetilde{\beta}}}(\Delta\Omega)^2\sin{\left(\epsilon\smash{+}\frac{\xi}{2}\right)},
\end{equation}
and its dependence on $\psi_0$ at $\Delta\Omega\smash{=}0.75$ is shown in Fig.\
\ref{su5}, whence it follows that the minimum of the spectral correlation function
lies near $\psi_0\smash{\approx}1$. In this case the photon antibunching takes place
for all phases $\psi_0\smash{>}0$ and it is maximal for $\psi_0\smash{\approx}1$. It
may be mentioned that the greater is the spectral band of measurement the stronger is
the photon bunching or antibunching.
\section*{DISCUSSION AND CONCLUSIONS}
The results presented in the present paper can be used for the correct interpretation
of the results of experiments \cite{Hashiura,Nishizawa,Rosenblug,Haus}, in which the
laser pulses with the duration of the order $100$ ps and quartz optical fibres were
used and the maximal meaning of nonlinear phase shift $\psi_{0}$ was greater than
$1$. Certainly, in the measurement of the quadrature spectrum the suppression of
quantum fluctuations of a pulse will be smoothed out (see (\ref{sist})). This time
over which the ``smoothing out" occurs in the case of balanced homodyne detection
\cite{Leonhardt} is determined by the duration of the heterodyne pulse.

The developed theory enables the choice of the optimal strategy at producing and
registration of ultrashort pulses in a squeezed states. The measurement of quantum
fluctuation of short pulses take place at high frequencies of the order of several
tens MHz in order to avoid any effects due to technical fluctuation concentrated at
low frequencies. However, in this area the suppression of quantum fluctuations is
greatest. The presented results show that by adjusting the phase of the signal pulse
(or the phase of a heterodyne pulse), maximal suppression of the quantum fluctuations
can be realized at the spectral component of interest for us. This spectral component
of interest can lie on the wing of the spectral response of nonlinearity (Fig.\
\ref{su2}). This means, that for obtaining squeezed-light pulses the nonlinear media
with a longer relaxation time and consequently with the greater nonlinearity can be
used \cite{Ahmanov}.

Our results suggest that in the spectral measurements the photon antibunching can be
observed. Usually, in the experiments spectral devices with confined spectral bands
are used, thus limiting the amplitude of vacuum fluctuations which participate in the
measurements. The final results show that the spectral correlation function depends
on the nonlinear phase addition, the relation between the pulse duration and the
relaxation time of the nonlinearity and also on the spectral band of the measurement.
In consequence the choice of the width of the spectral band of measurement can
represent an effective method of control of the photon bunching or antibunching. The
obtained results indicate that the photon antibunching can be observed at any value
of nonlinear phase addition in the low frequency measurements. At high frequency
measurements the photon bunching or antibunching strongly depends on the nonlinear
phase additions.

We note that the approach developed in the present article can be used to analyse the
formation of polarization-squeezed light in media with a cubic nonlinearity. This
will be treated in a future publication.
\section*{ACKNOWLEDGMENTS}
F.P. is grateful to S.~Codoban (JINR, Dubna) for useful discussions and rendered
help. The work has been performed with partial financial support from Programme
"Fundamental Metrology".

\end{document}